\documentclass[a4paper,10pt]{comjnl}
\usepackage{graphicx}
\usepackage{array}
\usepackage{multirow}
\usepackage{rotating}
\usepackage{listings}
\usepackage{amsmath}

\usepackage{color}

\begin{document}
\title[Accelerating cellular automata simulations using AVX and CUDA]{Accelerating cellular automata simulations using AVX and CUDA}
\author{Sebastian Szkoda}\email{sebastian.szkoda@ift.uni.wroc.pl}
\author{Zbigniew Koza}
\affiliation{Faculty of Physics and Astronomy, University of Wroc{\l}aw,  pl. M. Borna 9, Wroc{\l}aw,  Poland}

\author{Mateusz Tykierko}
\affiliation{Institute of Computer Engineering, Control and Robotics, Wroc{\l}aw University of Technology,  Wybrze\.ze Wyspia{\'n}skiego 27, 50-370 Wroclaw, Poland}

\shortauthors{S. Szkoda, Z. Koza, M. Tykierko}



\begin{abstract}
We investigated various methods of parallelization of the Frish-Hasslacher-Pomeau (FHP)
cellular automata algorithm for modeling fluid flow.
These methods include SSE, AVX, and POSIX Threads
for central processing units (CPUs) and CUDA for graphics processing units (GPUs).
We present implementation details of the FHP algorithm based on AVX/SSE and CUDA technologies.
We found that
(a) using AVX or SSE is necessary to fully utilize the potential of modern CPUs;
(b) CPUs and GPUs are comparable in terms of computational and economic efficiency only if the CPU code
uses AVX or SSE instructions; (c) AVX does not offer any substantial improvement relative to SSE.

\end{abstract}

\maketitle

\section{Introduction}

While for many years scientists and engineers could, to a large extend,
assume that each year their computer programs would run much faster only because of a steady, rapid growth of the hardware capabilities,
this assumption is no longer valid.
With processor clock speeds flattening or even dropping to save energy,
any further performance improvements must be coming from increased parallelism \cite{exascale}.
However, of the two major types of software parallelism, coarse-grained and fine-grained, so far only the former
has been widely adopted in  computational science,  mostly through  message-passing or shared memory interfaces,
whereas the latter tends to remain  a niche  technology, mostly because they introduce
an additional level of programming complexity.
This, however, leads to  underutilization  of the available resources in many areas of computational science,
a problem that is getting only more acute as new or improved hardware solutions supporting fine-grained
parallelism are being released.

Two such relatively new technologies are the advanced vector extensions (AVX),
supported by new generations of Intel and AMD processors,
and general-purpose computing on graphics processing units (GPGPU), supported by graphics cards from AMD and Nvidia.
Both  technologies exploit single instruction multiple data (SIMD) computing units for fine grained-parallelism.
However,
while AVX is a relatively simple extension to the i383 instruction set, GPU is a separate coprocessor
that must be programmed using dedicated programming languages and programming tools.
Although originally designed to support ``data-oriented'' programming in video and multimedia processing,
both can be used to accelerate general computations. This arises the question whether they can be used
to accelerate scientific computations on a much wider scale than to date.

Here we investigate AVX- and GPU-accelerated implementations of the Frish-Hasslacher-Pomeau (FHP) model of fluid flow
\cite{frisch86}. This model is an important example of a cellular automaton,
a broad class of numerical algorithms applicable in various areas of science and engineering \cite{wolfram02,chopard98}.
Typical cellular automata reduce the physical space to a regular grid of cells,
each one in a finite number of states,  evolving concurrently in discrete time steps
according to some rules based on the states of neighboring cells.
What distinguishes cellular automata from
other numerical models used in physics is that the states are represented by small integers
or even small sets of independent boolean variables rather than by floating-point numbers.
Thus, from a programmer's point of view,
FHP is an example of a numerical algorithm in which  huge amounts of integer data are arranged in a regular grid,
and the evolution can be performed in parallel  using bitwise
operators between the data fetched from the nearest-neighbor cells.
Note that even though this algorithm is amenable to massive fine-grained parallelization,
so far most research has focused on sequential architectures \cite{chopard98,kohring92}.
Therefore our goal is to investigate its parallelization properties.

Of the two general-purpose computing systems available for GPUs,
OpenCL (open standard) and CUDA (proprietary, Nvidia) \cite{Farber2011},
we chose the latter for its better support for scientific computing.
In contrast to CPUs, GPUs run in a multi-core mode by default.
Therefore a fair comparison between GPU  and CPU implementations requires that their efficiency
be compared processor for processor rather than core for core \cite{Debunking}.
Therefore in the CPU implementation we used POSIX Threads
to fully utilize all available processor resources.
We also implemented a SSE version of the CPU code to check
how much AVX outperforms its predecessor, SSE.
We believe that other cellular automata algorithms exhibit similar
potential for acceleration using AVX and GPU technology.

\section{Model}
\begin{figure}[h!]
\centering
\includegraphics[width=0.5\linewidth]{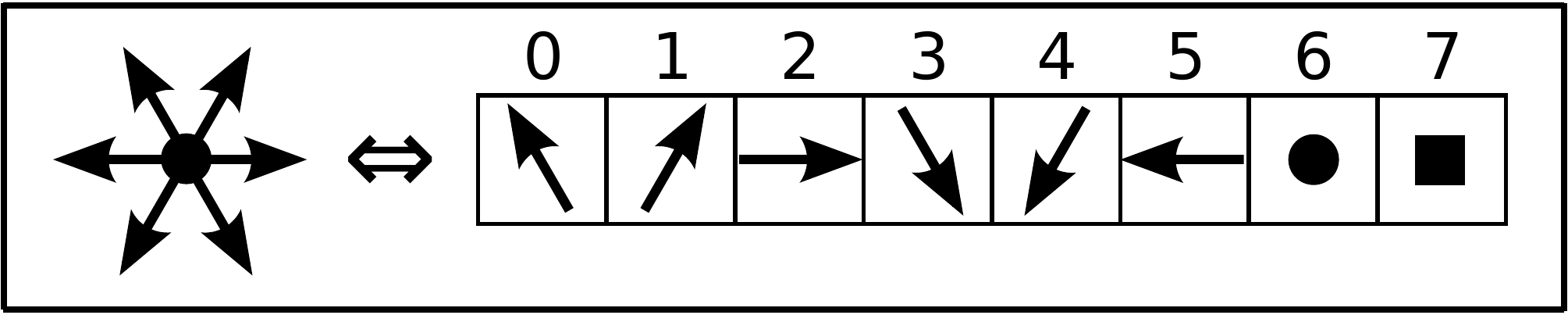}%
 \caption{The state of each node is represented by an 8-bit word. Bits 0-5 are mapped into particles with given nonzero velocities,
 bit 6 corresponds to a particle at rest, and bit 7 controls whether the node is a boundary node. \label{fig:bit}}
\end{figure}

In the FHP model one considers a triangular lattice
updated at discrete time steps $\Delta t$.
Seven boolean variables are assigned to each of its nodes
to represent the presence or absence of a particle with a given velocity $\mathbf{v}_i$, $i=0,\ldots,6$.
The values of velocities are related to the lattice vectors $\mathbf{c}_i$ through $\mathbf{v}_i = \mathbf{c}_i/\Delta t$, $i = 0,\dots, 5$,
and $\mathbf{v}_6 = 0$ corresponds to a particle at rest.
Thus, at each lattice node there can be up to 7 particles, each with a different velocity.
Since the FHP model is designed to simulate fluid flows in arbitrary (two-dimensional) geometries,
usually the 8-th boolean variable is added to each node to distinguish whether it is occupied by fluid or
by impenetrable boundary. Consequently, a convenient representation of a node state is an 8-bit word, or byte, see Fig.~\ref{fig:bit}.
For example, a bit pattern $00100100$ corresponds to a node with two particles,
one moving with velocity $\mathbf{v}_2$ and the other with $\mathbf{v}_5$, located in the area free occupied by the fluid.

\begin{figure}[h]
\centering%
\includegraphics[width=0.5\linewidth]{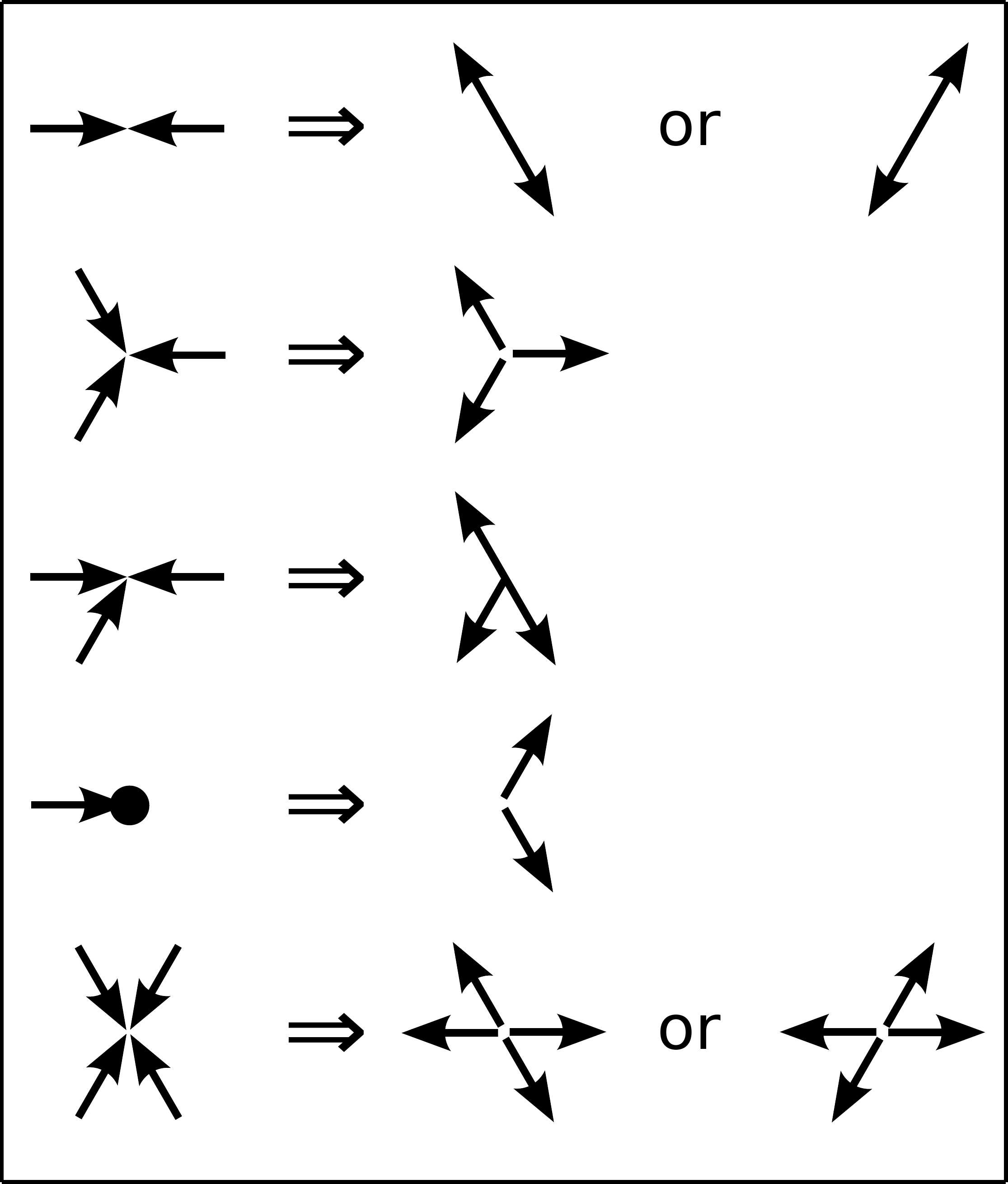}%
\caption{FHP collision rules.}
\label{fig:rules}%
\end{figure}

Evolution in each time step consists of two parts:
\begin{itemize}
 \item motion -- each particle moves to the adjacent node or stays at rest, according to its current velocity,
 \item scattering -- particles in each node collide and change their velocities
 according to the physical laws of conservation of momentum and mass,
 see Fig.~\ref{fig:rules}. This step includes bouncing from the boundaries.
\end{itemize}

\begin{figure}[h]
\centering
\includegraphics[width=0.7\linewidth]{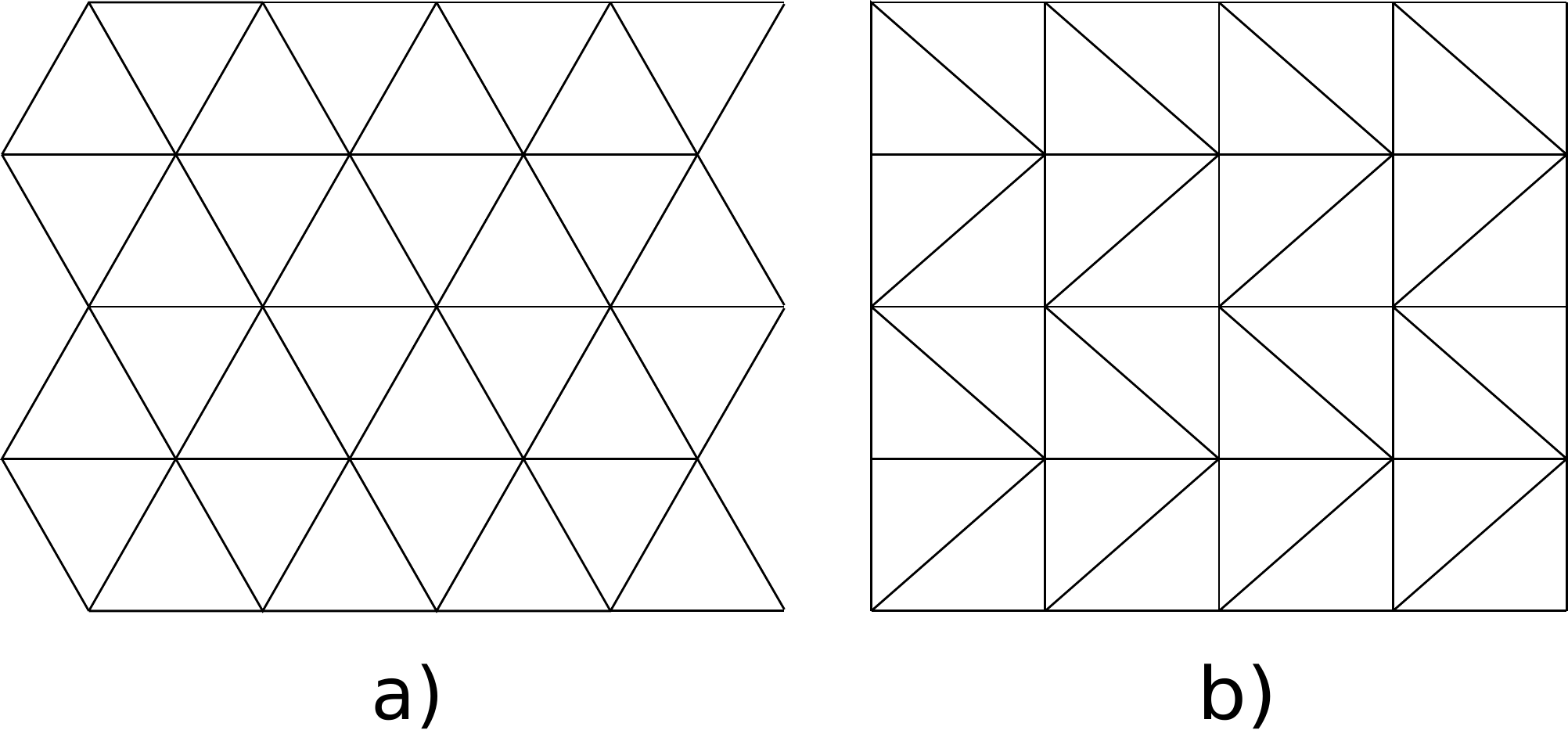}%
 \caption{ A triangular lattice (a) can be mapped onto a rectangular lattice by shifting
   every second row by half the lattice constant (b).
 \label{fig:lattice_mem}
}
\end{figure}

To implement the motion step we used two two-dimensional arrays holding the system state immediately before and after
translation of mobile particles. During this step, particles from the first array were propagated
to the appropriate neighboring nodes in the second array, according to their velocities.
Since the FHP is defined on a triangular lattice, we mapped its nodes into a rectangular
lattice, see Fig. \ref{fig:lattice_mem}. This lattice was then mapped into a two-dimensional computer array.

As soon as all particles have been propagated, the arrays are swapped and the collision step is performed.
Collision in each node is a local operation that does not involve neighboring nodes.
We implemented it using a look-up table, exploiting the fact that an 8-bit word can be regarded as
an integer in the range of $0,\ldots,255$.

\begin{figure}[h]
\centering%
\includegraphics[width=0.5\linewidth]{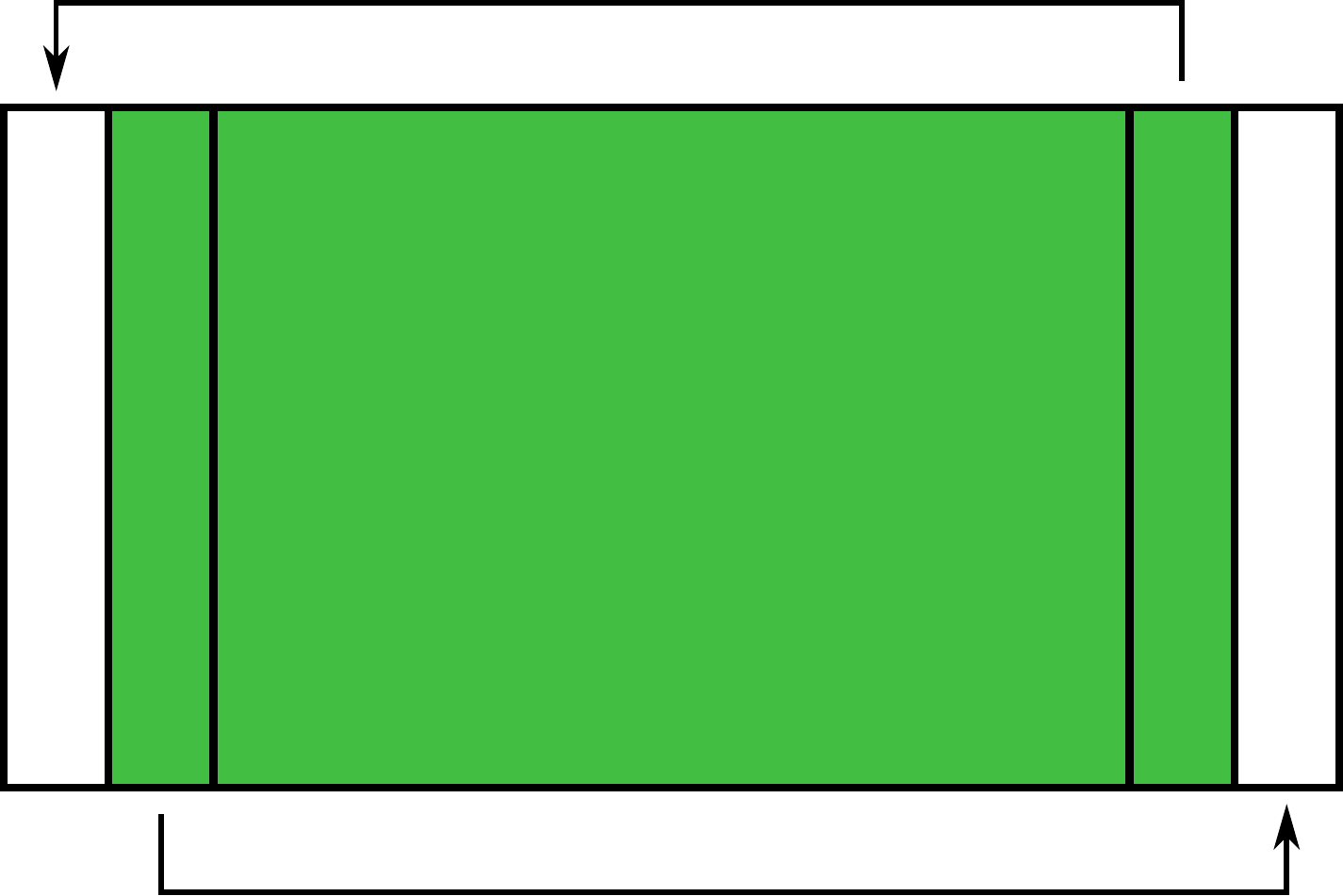}%
\caption{Schematic of the periodic boundary conditions. Each edge column is mirrored in an
extra column added on the other side of the grid.
}
\label{fig:BC}%
\end{figure}

Collisions with the boundaries
are defined through the the same look-up table as used for inter-particle collisions \cite{chopard98}.
We imposed reflecting (no-slip) boundary conditions on the top and bottom of the system and periodic boundary condition
in the horizontal direction.
Periodic boundary condition were implemented by extending the system by two columns, one on each side, each mirroring
the column on the opposite edge of the system, see~Fig.~\ref{fig:BC}.
In each step 
the first column is copied to the additional column on the right side,
and similarly the last one is copied to the extra column on the left.

To enforce fluid flow, we imposed an external constant force acting horizontally on the particles.
To this end, whenever the bits in a node fitted the pattern (..1..0..), where ``.'' denotes 0 or 1,
with some
probability $p$ we exchanged the bits on positions 2 and 5 to form a pattern (..0..1..).
This corresponds to reversing the velocity of a some of particle moving horizontally and results in the momentum transfer,
which, in turn, is equivalent to the action of an external driving a force.

\section{Implementations}

\subsection{CPU implementation with AVX}

\begin{figure}[h]
\centering%
\includegraphics[width=0.7\linewidth]{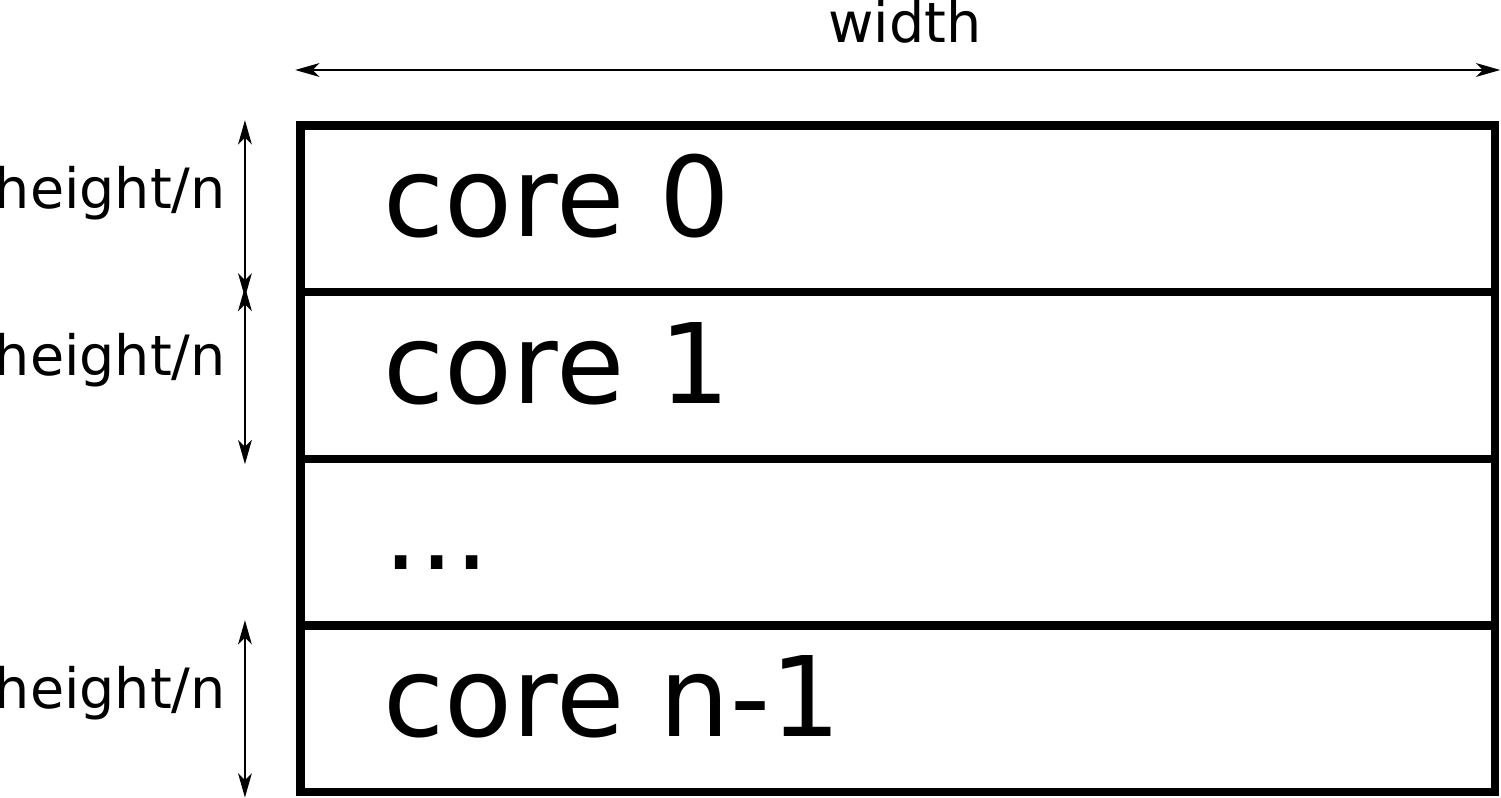}%
\caption{Grid division for POSIX Threads}
\label{fig:posix_div}%
\end{figure}

On the coarse-grained level the CPU implementation was parallelized using the POSIX Threads library.
The grid was divided into $n$ disjoint rectangular regions, where $n$ is the number o active threads,
see Fig.~\ref{fig:posix_div}.

At each time step  the threads must be synchronized
to prevent them doing the motion and scattering steps at the same time.
To this end each thread performs the motion step in its region, processing the rows from top to bottom and
writing the results to the auxiliary  array.
Synchronization is achieved by two barriers,
one just before the last row in the region is processed and the other one between
the motion and scattering steps. No data exchange between different threads is required.

Implementation of the motion step performed by individual threads was optimized with  AVX
to exploit the fact that 256-bit AVX registers allow to load and store the data
as well as execute instructions for $32$ nodes concurrently.
The details are shown in Listing \ref{code:avx}.
First, the data from 32 subsequent nodes are load into an AVX register.
Next, for $i=0,\ldots,5$ the implementation verifies whether the $i$-th bit at a given  node
is set and if so, sets the $i$-th bit in the appropriate neighboring node.
These tasks are achieved with boolean AND and OR operations performed in AVX registers,
taking into account that positions of neighbors of a given node depend on
whether this node belongs to an even or odd row, see Fig.~\ref{fig:lattice_mem}b.
Finally, positions of the particles at rest must also be updated.
Note that thanks to the AVX, all these actions are performed simultaneously on 32 lattice nodes.

\begin{lstlisting}[label={code:avx},
                   caption={The AVX kernel},
                   basicstyle=\small\ttfamily,
                   columns=fullflexible,
                   frame={tb}]
#define pos(i,j)  i*width + j
//loop over lattice nodes assigned for
//the current thread
for(j=start;j<end;j++)
{
  for(i=1; i<w-1;i+=32)
  {
    //read data
    valv = (__m256i*)(lattice0+pos(i,j));
    __m256i br = _mm256_loadu_si256(valv);

    //distribute data over neighboring nodes
    for(k=0;k<=6;k++)
    {
      maskv = _mm256_set1_epi8(mask1[k]);
      ar    = (__m256i)_mm256_and_pd((__m256d)br,\
              (__m256d)maskv);

      //determine neighbor positions
      e1 = i+e[k][j&1][0];
      e2 = j+e[k][j&1][1];

      valp = (__m256i*)(lattice1+pos(e1,e2));
      __m256i cr = _mm256_loadu_si256(valp);
      ar = (__m256i)_mm256_or_pd((__m256d)cr,\
           (__m256d)ar);

      //store data in destination nodes
      _mm256_storeu_si256(valp, ar);
    }
    //update rest particles
    br = (__m256i)_mm256_and_pd((__m256d)br,\
         (__m256d)maskv2);

    _mm256_storeu_si256(valv, br);
  }
}
\end{lstlisting}

The SSE implementation is very similar to the AVX one. The main difference is that
SSE uses 128-bit registers, so that only 16 FHP nodes can be processed at the same time by an SSE unit.

\subsection{GPU implementation with CUDA}
C for CUDA is a programming language from NVIDIA Corporation for programming their graphics cards.
It is a minimal extension to the standard C language that enables to run and synchronize thousands of GPU threads running in parallel.
In this programming model, a GPU device is a coprocessor running kernels asynchronously with the host CPU.
The main task of the CPU is to transfer data to the GPU, launch GPU computational kernels in a given order,
and read the results from the GPU.

The main difference in programming GPUs and CPUs is that the GPU exposes to the programmer
its hierarchical memory system. Thus, it is the programmer's responsibility to manage data allocation
in appropriate GPU memory pools, to manage transfers of the data within the same or between different memory pools,
to synchronize thread execution and their accesses to the memory, and to avoid race conditions or deadlocks.
This can be a quite nontrivial task, as different memory pools, such as registers, shared memory, caches, read-only buffers,
and the global memory differ significantly in their size, speed and connectivity to GPU resources.
It is in sharp contrast to the standard CPU programming model, where  the memory is assumed to be linear and uniform,
and the computational performance on the instruction level is managed in hardware.

GPU threads are arranged into blocks. Threads within a block can communicate with each other via fast, on-chip shared memory
and can be synchronized. However, as the order in which blocks are launched is undefined,
it is not possible to synchronize threads in different blocks. Therefore perhaps the single most important
step in GPU programming partitioning of the computational kernel into blocks of threads, assignment of
work to individual threads and management of inter-thread communication.

\begin{figure}[h]
  \centering
  \includegraphics[width=0.4\textwidth]{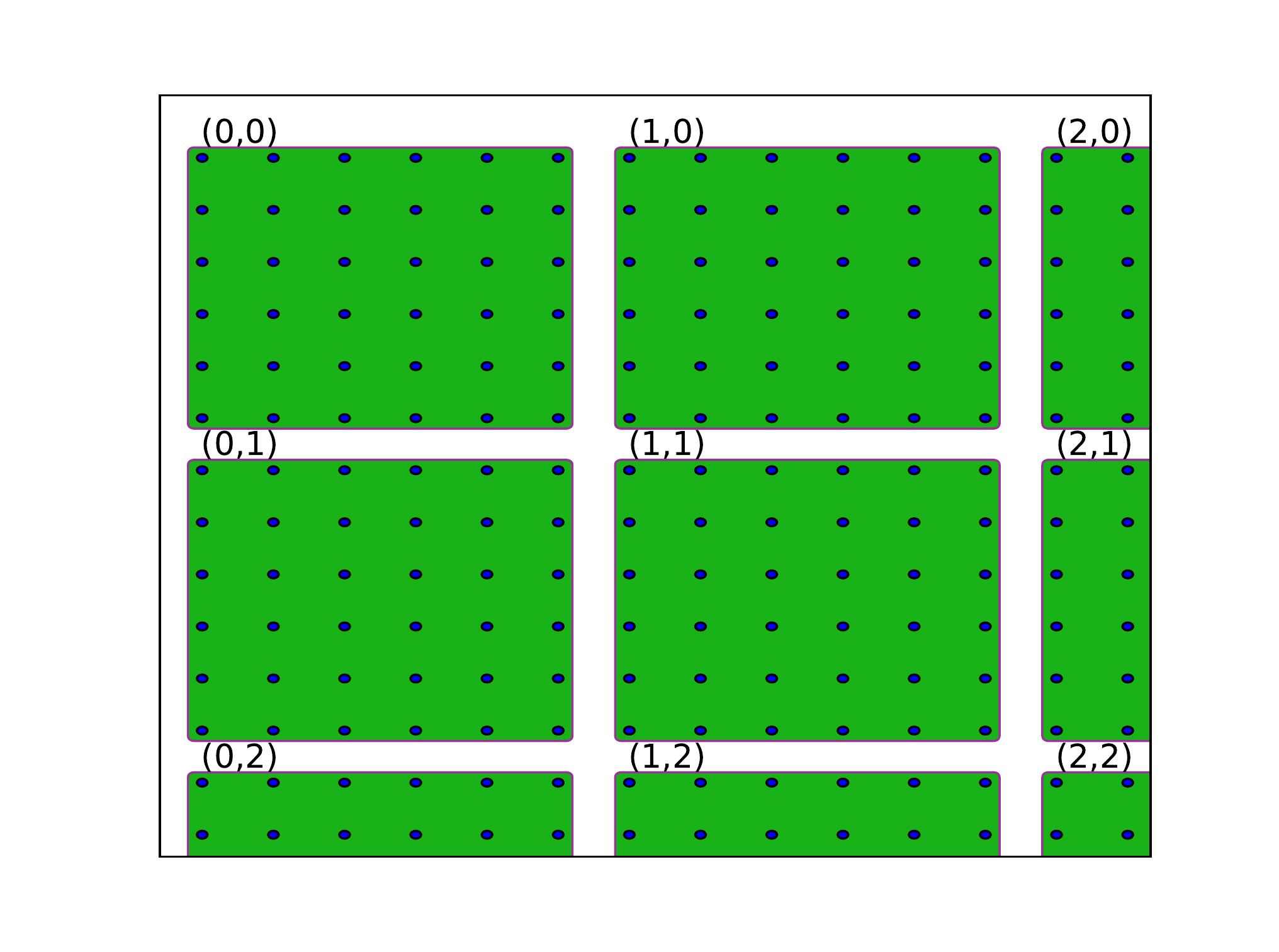}
  \caption{Naive division of the FHP lattice into computational blocks of nodes}
\label{fig:podzial1}
\end{figure}

The simplest approach consists in dividing the FHP lattice into disjoint rectangles, as shown in Fig.~\ref{fig:podzial1},
and assigning a node to a single thread.

This approach works well in the scattering step, because each node can be processed independently of each other
by applying the collision rules defined in the look-up table
and no communication between threads or blocks of threads is required.
There are also no memory conflicts, as the read and write operations to and from the global memory are executed only once per a node.

\begin{figure}
\centering%
\includegraphics[width=0.4\textwidth]{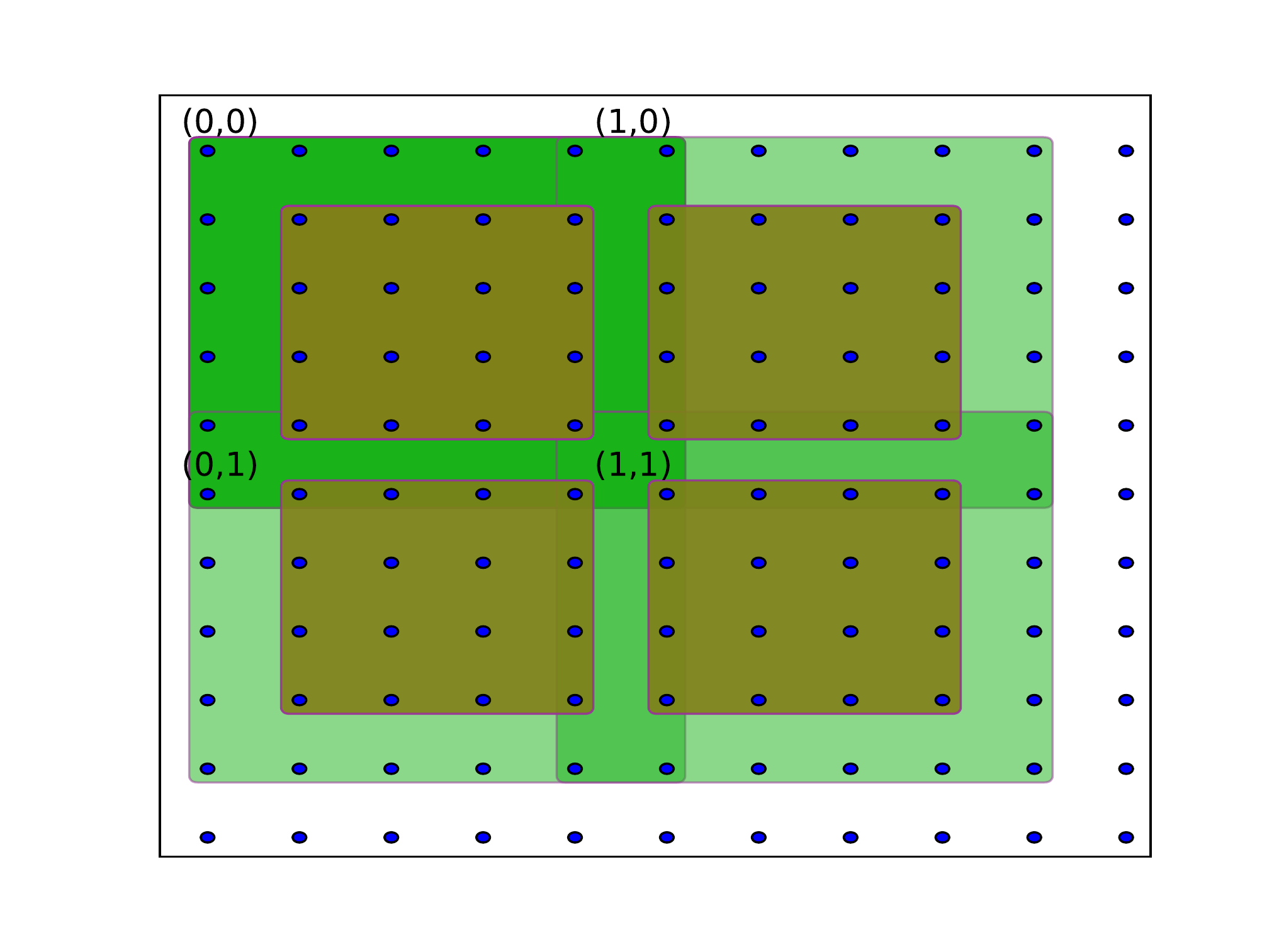}%
\caption{Partition of the FHP lattice into computational blocks of threads for efficient move step.
The size of the outer (green) rectangle is equal to the size of a block o threads.
This rectangle represents the nodes read by the corresponding block.
The inner (brown) rectangle marks the nodes whose state is updated ba a given block.
\label{fig:podzial2}}
\end{figure}

\begin{figure}
\centering%
\includegraphics[width=0.4\textwidth]{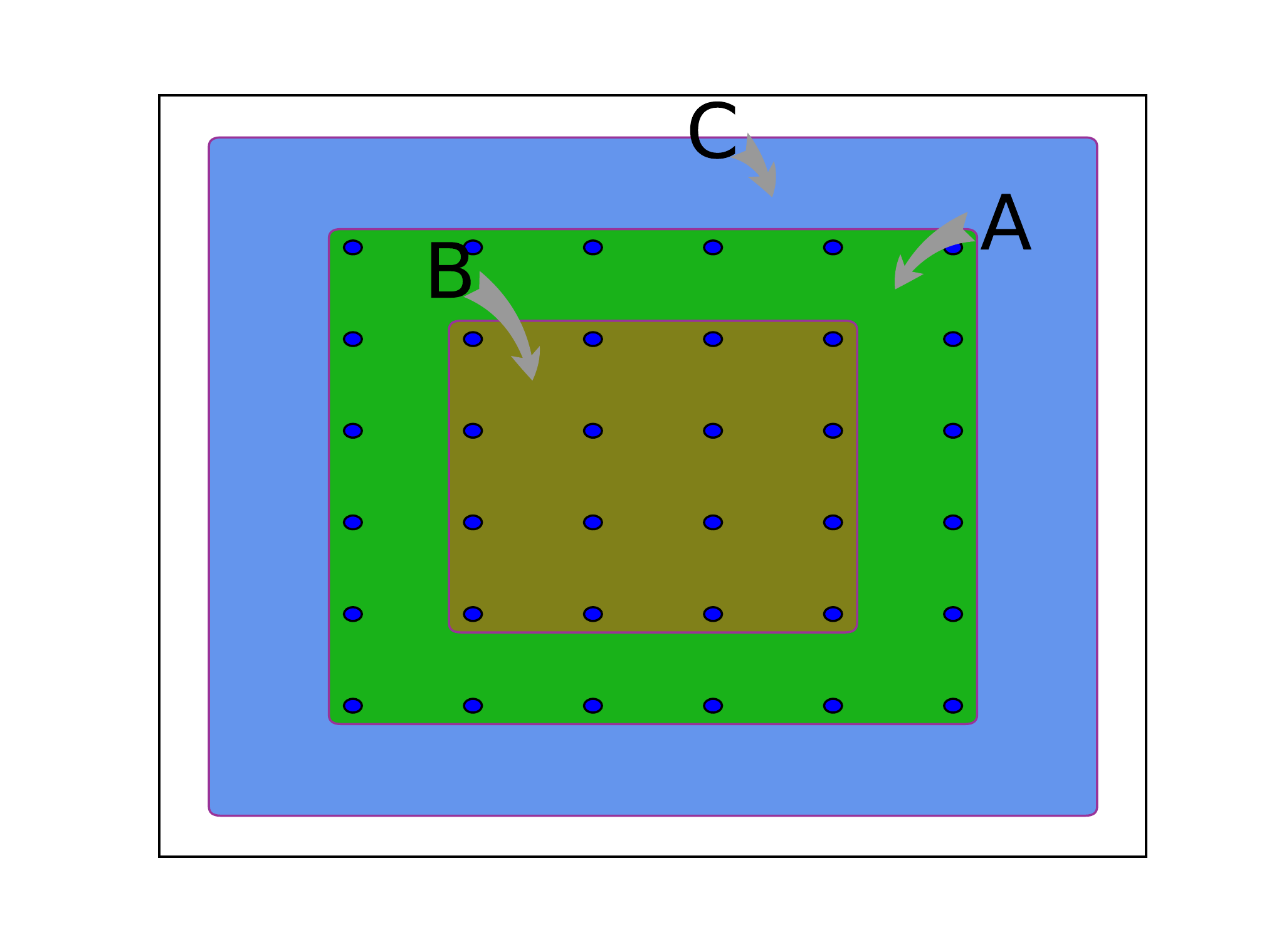}%
\caption
{
  Assignment of shared memory array elements to a block of threads.
  Rectangle A marks the FHP nodes loaded from the GPU main memory into the shared memory.
  Rectangle C embraces A and $C\setminus A$ is used as a buffer to capture the particles moving
  out of  A.
  Rectangle $B\subset A$ denotes the nodes whose updated states will be saved back to the main memory.
  Colors correspond to those in Fig.~\protect\ref{fig:podzial2}.
}
\label{fig:ABC}%
\end{figure}

However, such definition of computational blocks will not work well for the move step, as propagation of particles would require
communication and synchronization between different blocks. As this is impossible,
the only solution would be to us the global memory to store the data that would have to be communicated to neighboring blocks of threads
and use it in another kernel to complete the move step. Such a solution would be, however,
complicated and inefficient.
This problem can be solved by overlapping blocks, as depicted in Fig.~\ref{fig:podzial2}.
As illustrated in Fig.~\ref{fig:ABC}, a block of threads defines three rectangles, $A$, $B$ and $C$.
Rectangles marked as $B$ divide the lattice into disjoint parts, like in Fig.~\ref{fig:podzial1}.
They are embraced by rectangles $A$ corresponding to the blocks of threads.
Rectangles $A$, in turn, are embraced by rectangles $C$  corresponding to array elements in the shared memory.
Note that within this approach each FHP node is updated exactly once (rectangle $B$),
states of some nodes are read by 2, 3, or even 4 different blocks (rectangle $A$), and
the size of the array in the shared memory is larger than the block size (rectangle $C$).
Just as in the CPU implementation, a race condition is using different arrays for reading and writing the data.

\section{Results}
The AVX implementation was tested on  Intel i7~960 and two  server-class
Intel Xeon E3 and E5 processors.
The CUDA implementation was tested on server-class Nvidia Tesla M2075 computing devices and
on a high-end consumer graphics card, Nvidia GeForce GTX~480.
Since Xeon E5 was tested in a 4-processor configuration, we also investigated a system with two
GPUs connected by a PCI-Express switch. Also, we tested GPUs in two modes: one with error control and correction (ECC)
turned on and the other with ECC off. The latter mode is expected to yield $\approx$ 12\% faster access to the GPU memory.
The operating system was a 64-bit Linux and the CUDA version was 4.2.
The FHP lattice sizes used in simulations were restricted by the size of available memory
and amounted to $29,970\times27,990$ ($0.8 \times 10^9$ nodes) and
$46,850 \times 43,740$  ($2\times 10^9$ nodes) for the GPU and CPU versions, respectively.
The computational efficiency of each implementation was measured in million lattice site updates per second (Mups).
We also investigated the economic aspects of the simulations,
calculating the hardware purchase cost and electric power usage per Mups.
The cost of hardware components (in USD) was taken as a market price in July 2012.
For the GPUs the cost includes  the cost of purchasing or running corresponding CPUs.
The power consumption  was taken as the nominal values declared  by the producers.

The main results are shown in Table \ref{tab:tabelka}.
In subsequent columns it contains the information on the algorithm used, processor,
number of CPU cores used, number of CPU threads (which is twice the number of cores if the
Intel hyperthreading technology is used), computational efficiency in Mups, economic efficiency in USD/Mups and  Watt/Mups.

One thing that strikes in these results is that the extent to which the computational efficiency depends on the
parallelization method. Even if only one core is used, the program parallelized with SSE or AVX  can
run 3 times faster than its purely sequential counterpart.
Using all available cores usually gives even better acceleration, and the best results are obtained when
all parallelization methods are used. Particularly interesting conclusions can be derived from
the results for the Xeon E3 processor. While AVX can accelerate the sequential code by 3 times,
POSIX Threads on 4 cores and 8 threads accelerate the sequential code by 5.3 times, implying a superlinear dependency on the
number of cores. Interestingly, a similar superlinear behavior is observed for I7, for which
Hyperthreading was turned off. When combining Threads with AVX, the efficiency increases by 3.4
times relative to Threads-accelerated code and 18,2 times relative to the sequential code.
Interestingly, in all cases AVX overtakes its predecessor, SSE, by only $\approx 10\%$.
Note also that in all cases failing to use SSE or AVX results in underutilization of the computer resources by
a factor of about 2 to 3. An interesting aspect of the multi-GPU system its practically linear scaling with the number of GPUs.
This indicates that the FHP model can be efficiently simulated on larger GPU clusters.

\begin{table}\footnotesize
\begin{center}
\begin{tabular}{|c|c|r|r|r|r|r|}\hline
\rotatebox{80}{Code type}  &\rotatebox{80}{CPU type}   & \rotatebox{80}{CPU cores}
&\rotatebox{80}{CPU threads} &\rotatebox{80}{Mups}&\rotatebox{80}{USD/Mups} &\rotatebox{80}{Watt/Mups}\\\hline
seq        &Xeon i7         &1 &1 &28  &14.29&6.19\\
SSE        &960        &1 &1 &80  &3.33 &1.44\\
Pth        &(3.2GHz)   &4 &4 &138 &2.17 &0.94\\
pSSE       &           &4 &4 &340 &0.88 &0.38\\\hline
seq        &Xeon E3         &1 &1 &36  &9.17 &2.14\\
SSE        &1270       &1 &1 &96  &3.44 &0.80\\
AVX        &(3.5 GHz)  &1 &1 &108 &3.06 &0.77\\
Pth        &           &4 &8 &190 &1.74 &0.41\\
pSSE       &           &4 &8 &596 &0.55 &0.13\\
pAVX       &           &4 &8 &654 &0.50 &0.12\\\hline
SSE        &Xeon E5         &1 &1 &72  &50.00&1.60\\
AVX        &4650L      &1 &1 &80  &45.00&1.44\\
pSSE       &(2.6 GHz)  &32&32&1446& 9.96&0.31\\
pAVX       &           &16&16&1313& 5.48&0.17\\\hline
GPU        &GTX  480   &1 &1 &1746& 0.46&0.30\\
mGPU       &           &2 &2 &3493& 0.37&0.27\\\hline
GPU        &Tesla M2075     &1 &1 &1103& 2.72&0.32\\
GPU$^\dagger$          &      &1 &1 &1198& 2.50&0.30\\\hline
\end{tabular}
\end{center}
\caption{Main results.
\scriptsize {\newline Glossary:
   \newline \textbf{seq} -- sequential,
   \newline \textbf{SSE} -- Streaming SIMD Extensions,
   \newline	\textbf{AVX} -- Advanced Vector Extensions,
   \newline \textbf{Pth} -- POSIX Threads (PThreads),
   \newline \textbf{pSSE} -- PThreads with SSE,
   \newline \textbf{pAVX} -- PThreads with AVX,
   \newline \textbf{GPU} -- Graphics Processing Unit,
   \newline \textbf{mGPU} -- Multiple GPUs,
   \newline \textbf{GPU$\mathbf{^\dagger}$} -- a GPU model with ECC disabled.
   \label{tab:tabelka}
} }
\end{table}

The highest computational efficiency was found for a high-end gaming GPU, GTX480.
This is related to the fact that it is equipped with the fastest memory, which is the key
factor in the FHP modeling, since FHP is a memory-bound algorithm. However, the difference between
GTX480 and Xeon E5 is only 20\%, and the latter platform turns out even faster than the server-class Tesla M2075.
However, one should take into account that Xeon E5 is an 8-core processor and we tested it on 2- and 4-processor systems.
A processor-for-processor comparison reveals superiority of GPUs, with a single M2075 being about 2 times faster than a single Xeon E5.
This value is in accord with other reports on the acceleration a GPU can deliver over well-tuned CPU code \cite{Debunking}.

\begin{figure}
 \centering
  \includegraphics[width=0.4\textwidth]{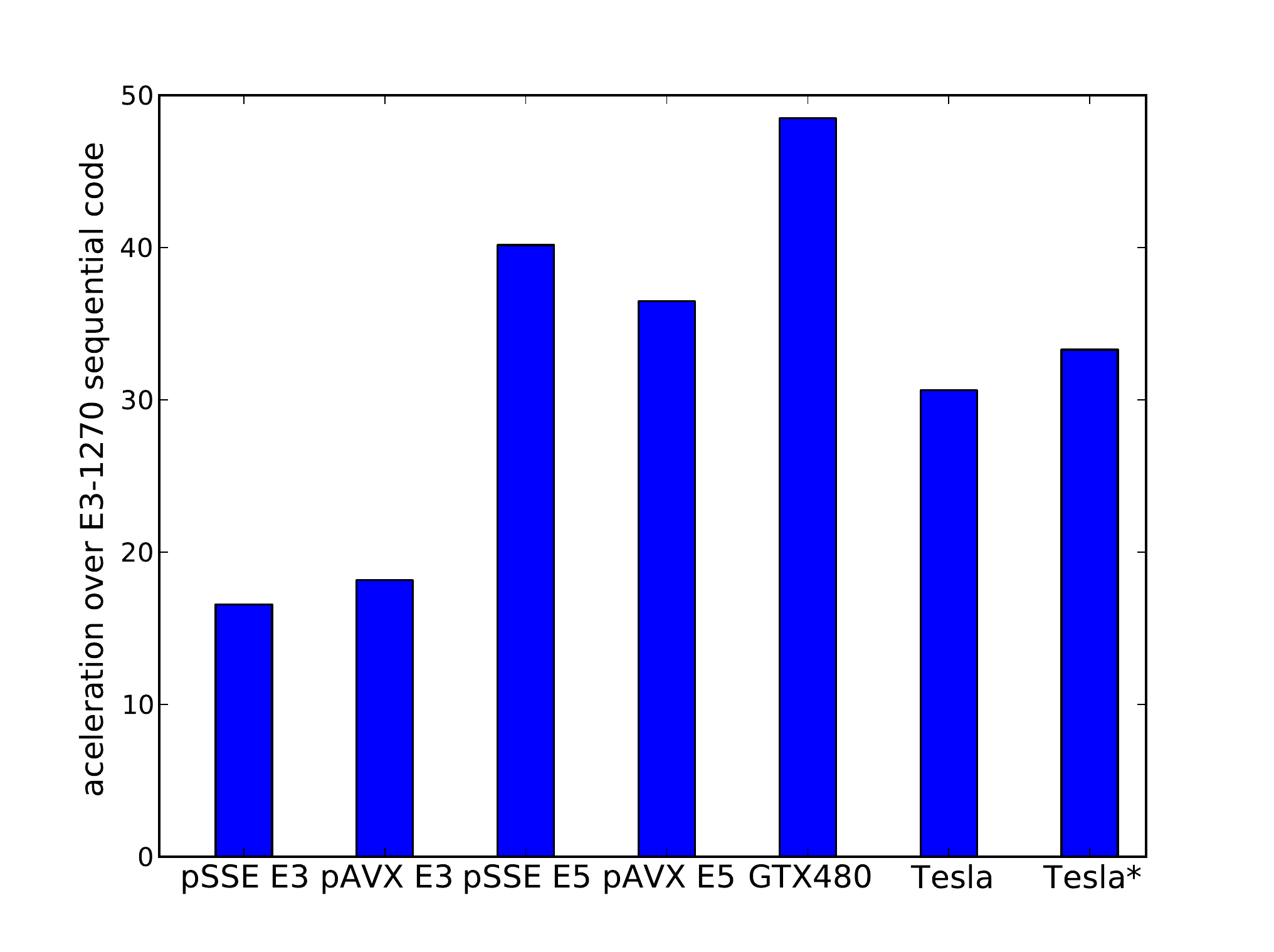}%
  \caption{Acceleration over sequential implementation on E3-1270 CPU.
            \label{fig:histogram2}
           }
\end{figure}

Figure \ref{fig:histogram2} shows the acceleration obtained on various processors using different
parallelization techniques.
The highest value, almost 50, is found for the GTX~480 graphics card. Note however,
that a well-optimized code on the same CPU platform is almost 20 times faster than the sequential code,
and the acceleration on the E5 reaches 40 times. Thus, it is not the architecture
of the GPUs that results in reports of it being tens, hundreds or thousands of times faster than the CPUs,
but the fact that this architecture enforces from the very beginning
 a very high level of parallelization on all levels.
In a code utilizing all ways of CPU parallelization, the GPU over CPU speedup drops below 10, in accordance
with the hardware capabilities of the devices under comparison.

\begin{figure}[h]
\centering
\includegraphics[width=0.4\textwidth]{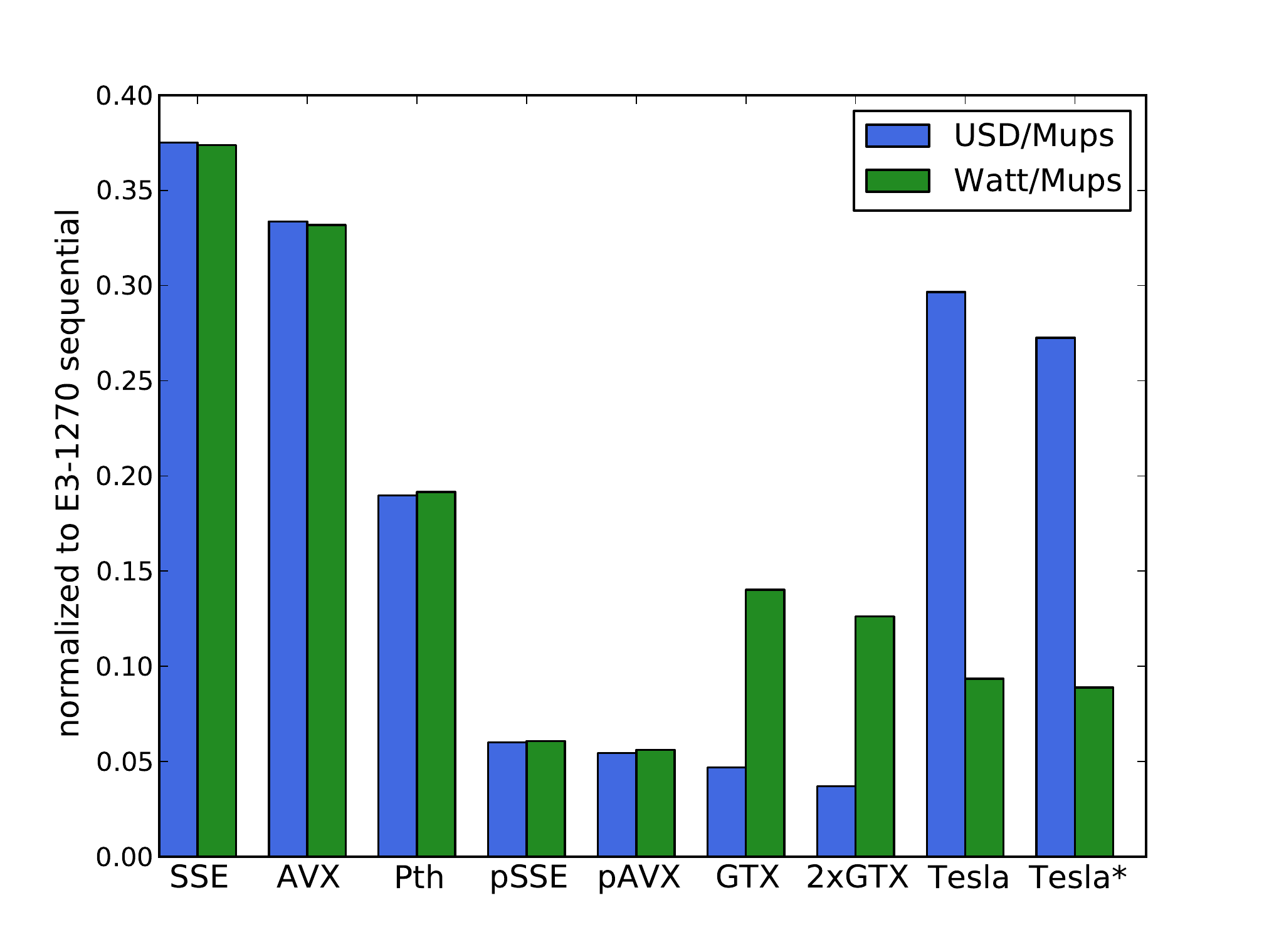}
 \caption{Comparison of the purchase (blue) and running (green) costs relative
          to the sequential implementation on the Xeon E3-1270 processor.
          All CPU data are for the E3 processor.
          \label{fig:histogram1}
         }
\end{figure}

Results for the economic aspects of running FHP simulations on different platforms are shown in Fig.~\ref{fig:histogram1}.
It turns out that the most cost and power efficient processor is E3, but only if the program
running on it was parallelized with the SSE or AVX technology.

\section{Conclusions and outlook}

We investigated several methods of fine-grained parallelization of the FHP cellular automata model.
We found that using SSE or AVX vector instructions in the CPU code can accelerate a single-thread program
by a factor of about 3, and combining this technique with POSIX Threads yields acceleration from 12 to 18 times.

While the most computationally efficient solution was found to run the simulation on GPUs, which turned out
up to about 2 times faster than the most efficient single-processor CPU, CPUs turned out cheaper to buy and cheaper to run,
but only if they exploited the  SSE or AVX technology. However, our results are based
on some \emph{ad hoc} assumptions, so that they cannot be used to decide which architecture is superior
for cellular automata simulations. First,
it is possible that each of our implementations could be further improved. Second,
calculation of the electric power efficiency for the GPUs includes the power drawn by the CPU and one could argue that
this overestimates real costs, as only one CPU core is occupied by the GPU, and up to 8 GPUs can be attached to a single
CPU using PCI-Express switches. 
Third, while our tests included one of the newest intel processors, Xeon E5, the GPUs
used in this study belong to a relatively old Fermi architecture and one can expect that
new GPU processors, especially Tesla K10, should reduce the power consumption 
by at least a factor of 2. 

In spite of these difficulties we can conclude that as far as cellular automata simulations are concerned:
(a) using AVX or SSE is necessary to fully utilize the potential of modern CPUs; (b) CPUs and GPUs are comparable in terms of
computational and economic efficiency; and (c) AVX does not offer any substantial improvement relative to SSE.

\section{Acknowledgments}
SS prepared this publication as part of the PRACE 2IP project receiving funding from the
EU's Seventh Framework Programme (FP7/2007-2013).
ZK acknowledges support from Polish Ministry of Science and Higher Education Grant No. N N519 437939.
Calculations have been carried out in Wroclaw Centre for Networking and Supercomputing (http://www.wcss.wroc.pl).

\end{document}